\documentclass[twocolumn,apjl,numberedappendix]{openjournal}

\usepackage{makecell}
\usepackage{fixes}
\usepackage{graphicx}
\usepackage{tikz}
\usepackage{xspace}
\usepackage{dutchcal}

\usepackage{tikzscale}          
\usepackage[english]{babel}
\usepackage{comment}

\newcommand{\twodsquaremat}[4]{\left( \begin{array}{cc}#1 & #2\\ #3 & #4 \end{array} \right)}

\definecolor{APSblue}{HTML}{2E3092} 
\usepackage[colorlinks=true, linkcolor=APSblue, urlcolor=APSblue, citecolor=APSblue]{hyperref}
\usepackage{comment}
\usepackage{orcidlink}
\usepackage{fontawesome5}
\usepackage{amsmath}
\usepackage{esint}
\usepackage{mathrsfs}
\usepackage[capitalise]{cleveref}

\begin{document}

\title{Calibration of CMB Polarisation Using Cross-Experiment Correlations}
   \author{%
Claire Rigouzzo\textsuperscript{1}\,\orcidlink{0000-0002-1551-4018},
Eugene Lim\textsuperscript{1}\,\orcidlink{0000-0002-6227-9540},
Susanna Azzoni\textsuperscript{2}\,\orcidlink{0000-0002-8132-4896},
Yiqi Liu\textsuperscript{2}\,\orcidlink{0000-0002-5210-8035}%
}
\email{claire.rigouzzo@kcl.ac.uk}

\affiliation{%
\textsuperscript{1}Laboratory for Theoretical Particle Physics
and Cosmology, King's College London, London, United Kingdom\\
\textsuperscript{2}Joseph Henry Laboratories of Physics,
Princeton University, Princeton, NJ 08544, USA
}
\begin{abstract}
Parity-violating physics in the Universe can generate correlations between the Cosmic Microwave Background (CMB) $E$- and $B$-modes, but detecting such signals requires extremely accurate calibration of instruments. We describe a data-driven method to calibrate the relative polarisation angle between CMB experiments using cross-correlations of observations over a common sky region. Unlike standard self-calibration approaches, this method does not assume vanishing isotropic cosmic birefringence or primordial $EB$ correlations when estimating the relative misalignment angle, and therefore preserves sensitivity to parity-violating physics. As a proof of concept, we forecast the performance of this method using the Simons Observatory (SO) Small Aperture Telescopes (SATs) as a calibrated reference. If they can be calibrated to an uncertainty of $0.08^\circ$, as anticipated from the SO wire grid calibration system, we show that the SO Large Aperture Telescope and \textit{Planck} could be calibrated to uncertainties of $0.10^\circ$ and $0.17^\circ$, respectively, at $\sim 145$ GHz.
This approach relies on the availability of at least one well-calibrated instrument, and provides a complementary path to improving polarisation calibration across experiments, enabling more robust searches for parity-violating physics in the CMB, such as cosmic birefringence.
\end{abstract}

\maketitle
\section*{Introduction}
The Cosmic Microwave Background (CMB) is a central pillar of precision cosmology, providing a uniquely clean probe of the physics of the early Universe and the growth of structure \cite{1965ApJ142414D,Peebles:1968ja}. In addition to temperature anisotropies \cite{Spergel_2003,Das_2011,Keisler_2011,theplanckcollaboration2006scientific}, the CMB exhibits linear polarisation due to Thomson scattering at the last scattering surface \cite{Komatsu_2011,1968ApJ153L1R,10.1093/mnras/191.2.207,10.1093/mnras/202.4.1169}. This signal is described by the Stokes parameters $Q$ and $U$ \cite{jackson_classical_1999,2003ApJS148161K}, typically recast into parity eigenmodes $E$ and $B$ \cite{Kamionkowski_1997,Kosowsky_1996,PhysRevLett.78.2054}.

Polarisation signals are much fainter than temperature fluctuations: $E$-modes are $10$–$100$ times weaker, and $B$-modes are further suppressed. Furthermore, unlike temperature, polarisation transforms under rotation, making it highly sensitive to instrumental misalignments. Accurate calibration is therefore essential to recover the $E$- and $B$-modes precisely. Thermo-mechanical deformations of the receiver structure can lead to polarisation angle misalignments at the level of $0.1^\circ$, even before accounting for additional systematics such as differential beam effects or magnetic pickup \cite{Planck_2016_systematic}. Even small angular errors leak $E$-modes into $B$-modes \cite{PhysRevD.67.043004,Shimon:2007au}, severely limiting our ability to detect a tensor-to-scalar ratio $r \sim \mathcal{O}(10^{-2})$ \cite{Keating_2012} — a key target for upcoming polarisation experiments. 

Instrumental misalignment induces spurious $TB $ and $ EB $ correlations, that are observationally indistinguishable from parity-violating signals such as isotropic cosmic birefringence \cite{Komatsu_2022}. A true detection of $EB$ correlation would indicate new physics such as Lorentz violation \cite{Colladay_1998,Kosteleck__2002,Kosteleck__2007}, axion-like fields along the line of sight \cite{Harari:1992ea,Komatsu_2022}, or inflationary parity violation \cite{Anber_2010,Sorbo_2011}. Separating such effects from instrumental systematics remains a major challenge in the search for physics beyond the Standard Model \cite{Carroll_1998,Lue_1999,Feng_2005}. 

Here, we propose to exploit cross-correlations between experiments as a direct estimator of relative polarisation-angle misalignment, in a way that preserves sensitivity to parity-violating signals. A related strategy was explored by BICEP/Keck in \cite{BICEPKeck:2024cmk}, where CMB cross-spectra are used to infer relative polarisation-angle offsets. In the present work, we focus on explicitly propagating the residual uncertainty in this relative calibration into the systematic error budget for parity-violation measurements. Unlike self-calibration approaches, which impose vanishing $EB$ correlations by construction, this method does not assume the absence of parity-violating signals, and therefore preserves sensitivity to cosmic birefringence. The method determines relative misalignment angles and is anchored by an independently calibrated reference instrument. In this work, we adopt the Simons Observatory (SO) Small Aperture Telescopes (SATs) as this reference.

As a concrete application, we demonstrate how this approach can leverage the expected precise calibration of the SO SATs \cite{Murata_2023,SO:2024ntl,2024ApJ961138D,2024SPIE13102E19C} to calibrate other telescopes. The Simons Observatory, located in Chile, consists of two complementary surveys using the 
SATs and the Large Aperture Telescope (LAT) 
\cite{SimonsObservatory:2025wwn}. These instruments serve distinct science goals. The SAT survey covers approximately $10\%$ of the sky and is optimized for precise measurements of $B$-mode polarisation at larger angular scales ($30 \lesssim \ell \lesssim 300$). In contrast, the LAT survey targets about $40\%$ of the sky, with substantial overlap with large-scale structure experiments such 
as DES \cite{DES:2026fyc}, DESI \cite{DESI:2025zgx}, and Rubin \cite{LSST:2008ijt}. The SAT science goals require dedicated, high-precision absolute polarisation calibration campaigns. We will show that the resulting calibration accuracy can also be leveraged by the LAT and other CMB experiments observing overlapping regions of the sky.

We describe how this method can be used to calibrate other instruments via cross correlation with the SATs, specifically for \textit{Planck} \cite{Planck:2016soo,Planck:2018gnk} and LAT data. 
This cross-calibration would tighten constraints on isotropic polarisation 
misalignment and provide a clearer window onto parity-violating physics beyond 
the Standard Model.  

The paper is organized as follows. In Section \ref{sec:prob_misca}, we introduce the problem of miscalibration and its impact on the $E$- and $B$-modes. In Section \ref{sec:parity_calib}, we examine how different parity-breaking mechanisms imprint signatures on the observed $EB$ power spectrum, and we review existing calibration strategies, including their precision and underlying assumptions. In Section \ref{sec:method}, we present a calibration method that uses cross-correlations between different instruments, and assess its performance, first analytically and then through tests of its robustness against simulations. We next illustrate how this method can be used to improve constraints on parity-breaking physics in Section \ref{sec:parity}. Finally, in Section \ref{sec:discussion}, we discuss the advantages and limitations of the method, along with directions for future improvement.

\paragraph*{Conventions.}
Throughout, we quote uncertainties as the $68 \%$ C.L.
\section{The problem of Miscalibration}
\label{sec:prob_misca}
Accurate recovery of the CMB $E$- and $B$-mode signals requires precise knowledge of the instrument’s polarisation orientation. A small miscalibration in the polarimeter orientation rotates the measured polarisation basis and induces leakage from $E$- to $B$-modes \cite{PhysRevD.67.043004,Shimon:2007au,Keating_2012}. In such cases, the actual orientation angle $\psi$ of the polarimeter differs from the intended design angle $\psi_\text{design}$ by a misalignment angle $\alpha$:
\begin{equation}
\psi = \psi_\text{design} + \alpha \; .
\label{def_alpha}
\end{equation}
Expressed in terms of spherical harmonic modes, the resulting mode mixing takes the form \cite{Zhao_2015,Minami:2019ruj,Diego_Palazuelos_2023,Komatsu_2022}:
\begin{equation}
\begin{aligned}
E^o_{\ell m} &= \cos (2\alpha) E_{\ell m} - \sin (2\alpha) B_{\ell m} , \\
B^o_{\ell m} &= \sin (2\alpha) E_{\ell m} + \cos (2\alpha) B_{\ell m} ,
\end{aligned}
\label{leakage}
\end{equation}
where $E_{\ell m}$ and $B_{\ell m}$ denote the \emph{true} sky modes, while $E^o_{\ell m}$ and $B^o_{\ell m}$ represent the \emph{observed} modes affected by the misalignment angle $\alpha$. Throughout this work, the superscript ``o'' designates observable quantities, while unscripted symbols refer to the corresponding true sky values. The observed $EB$ power spectrum therefore reads \cite{Zhao_2015,Gruppuso_2016}:
\begin{equation}
    C_\ell^{EB,o} = C_\ell^{EB} \cos(4\alpha) 
    + \left(C_\ell^{EE} - C_\ell^{BB}\right) \cos(2\alpha)\sin(2\alpha) \; . 
    \label{E_o_B_o}
\end{equation}
The observed $C_{\ell}^{E B,o}$ thus receives two contributions: a sky term $C_{\ell}^{E B}$, and a rotation-induced term proportional to $C_{\ell}^{E E}-C_{\ell}^{B B}$. In practice, the observed rotation may arise from both sky-originating signals – cosmological (e.g., cosmic birefringence) or astrophysical foregrounds – and instrumental polarisation angle miscalibration.

Our goal is to disentangle the contribution arising from the misalignment angle from the true sky signal $C_{\ell}^{E B}$, which encodes information about parity violation. Note that we do not attempt to separate a cosmological contribution from potential foreground-induced $E B$ correlations. Instead, we assume negligible $EB$ contribution from the Galactic foreground at the nominal observing frequencies considered in this work (~$\sim~90-150~\mathrm{GHz}$~). However, this assumption may not hold for real observations \cite{Diego-Palazuelos:2022cnh,Hervias-Caimapo:2024ili}, and a more complete analysis, including foreground-induced parity-violating signals, is left for future work. An upcoming Simons Observatory paper will present a more detailed treatment of the SAT–LAT cross-calibration, with particular emphasis on foregrounds~\cite{lonappancoming}. In the following, we present two main sources of parity violation and we examine how this degeneracy might be disentangled to reliably probe parity-breaking physics.

\section{Parity Breaking and Calibration}
\label{sec:parity_calib}
There are several mechanisms by which parity-violating physics could manifest in the CMB. One possibility is that the cosmological background through which CMB photons propagate breaks parity symmetry, for example if a pseudoscalar field couples to photons via a Chern-Simons-like term \cite{Harari:1992ea,Komatsu_2022}. Another possibility is that parity-violating effects arise already at the surface of last scattering. Examples include chiral gravity, axion-like inflation, and related models \cite{Anber_2010,Sorbo_2011,Barnaby_2012,Maleknejad_2011,Maleknejad:2011jw,Alexander_2009,Saito_2007,Contaldi:2008yz}. We will briefly review these mechanisms and discuss whether current calibration methods are sensitive to such effects.
\subsection{Isotropic Cosmic Birefringence}
The presence of parity-violating interactions for photons can be measured via the rotation of their polarisation plane. Consider for example a pseudoscalar field $\phi$, that couples to photons through a Chern-Simons term $\phi F_{\mu \nu} \tilde{F}^{\mu \nu}$, where $\tilde{F}_{\mu \nu}= 1/2 \epsilon_{\mu \nu \alpha \beta } F^{\alpha \beta}$. As is shown in Appendix~\ref{iso_birefringence}, the presence of such a term would impact the right-handed and left-handed photons asymmetrically, rotating the plane of polarisation between the last scattering surface and today. The amount of rotation is characterised by the cosmic birefringence angle $\beta$ (see \cite{Komatsu_2022} for a review). Note that such a rotation due to parity breaking can be generated by more general models such as CPT or Lorentz breaking \cite{Colladay_1998,Kosteleck__2002,Kosteleck__2007}.

The main challenge is that an isotropic cosmic birefringence angle $\beta$ is completely degenerate with instrumental systematics. Specifically, a rotation $\beta$ from parity-violating physics is observationally indistinguishable from a telescope misalignment $\alpha$. The total rotation of the polarisation plane $\tilde{\alpha}$ is simply
\begin{equation}
    \tilde{\alpha} = \alpha + \beta \, ,
    \label{alpha_total}
\end{equation}
where $\alpha$ denotes instrumental misalignment and $\beta$ the cosmic birefringence. \emph{In practice, $\alpha$ must therefore be replaced by $\tilde{\alpha}$ in Eqs.~(\ref{leakage})–(\ref{E_o_B_o})}. This degeneracy significantly limits our ability to constrain $\beta$, which would otherwise serve as a smoking-gun signal of parity-violating cosmology \cite{Komatsu_2022}.

While the telescope misalignment $\alpha$ is expected to be isotropic, cosmic birefringence may also contain an anisotropic component: $\beta = \beta_0 + \beta(\hat{n})$. The isotropic rotation $\beta_0$ corresponds to a uniform polarisation rotation across the sky, typically sourced by a homogeneous background field or a constant coupling to a pseudoscalar field \cite{Carroll:1989vb,Lue:1998mq}. In contrast, the anisotropic rotation $\beta(\hat{n})$ is generated by spatial fluctuations in such fields or by line-of-sight variations in couplings, and manifests as a direction-dependent polarisation rotation. The anisotropic $\beta(\hat{n})$ can be probed with quadratic estimators and cross-correlations \cite{Kamionkowski_1997}, and current analyses find no significant detection \cite{Bianchini_2020,Namikawa_2020,Gruppuso_2020,Jain_2022}. By comparison, the isotropic component is much harder to constrain: it is nearly degenerate with instrumental angle  miscalibration $\alpha$ \cite{Takahashi_2010,Keating_2012,Bischoff_2013,Komatsu_2011}, and isolating it requires absolute calibration methods. From here on, we focus exclusively on this isotropic component.
\subsection{Primordial EB spectrum}
Another potential signature of parity-violating physics may originate at (or before) the 
surface of last scattering. Consider a pseudoscalar field $\phi$ that either 
drives inflation \cite{Anber_2010,Sorbo_2011} or exists as a light spectator 
field, that is, a subdominant degree of freedom that does not control the 
background expansion but can still couple to gauge fields. In such scenarios, 
a Chern--Simons coupling can induce parity violation in the early universe and 
leave observable imprints on primordial perturbations \cite{Barnaby_2012}. Such interactions can generate chiral primordial gravitational waves 
\cite{Sorbo_2011,Barnaby_2011}, leading to a non-zero primordial $EB$ power 
spectrum. Similar parity-violating signatures may also arise from gauge field 
production mechanisms \cite{Maleknejad_2011,Maleknejad:2011jw} or within chiral 
gravity models \cite{Alexander_2009,Saito_2007,Contaldi:2008yz}.

As shown in \cref{E_o_B_o}, an observed $EB$ power spectrum may arise from a primordial $EB$ signal, or from a uniform polarisation rotation due to a combination of instrumental misalignment $\alpha$ and isotropic cosmic birefringence $\beta$. If the primordial $EB$ spectrum has a well-motivated shape, for example, in chiral gravity \cite{Gluscevic_2010} or axion–$SU(2)$ inflation \cite{Thorne:2017jft}, it can in principle be separated from the rotation contribution. However, at the level of two‑point functions there is an exact degeneracy between a uniform primordial $EB$ signal and a uniform rotation. Disentangling the two requires external calibration or additional assumptions (e.g., foreground‑based multi‑frequency methods or model‑specific $EB/TB$ templates).

\subsection{Calibration}
We have seen that parity-breaking effects manifest themselves through a non-zero observed $EB$ correlation. We now summarize the main existing calibration strategies, highlighting the assumptions they make. 
\paragraph*{Self-calibration:} Assuming that there is no isotropic birefringence $\beta$ and no primordial $EB$ power spectrum, one can solve for $\alpha$ directly from the measured $TB$ and $EB$ spectra \cite{Keating_2012}. This method achieves a very good precision ($\sim 0.03^\circ$ \cite{2016_parity,AtacamaCosmologyTelescope:2025blo}), but it is fundamentally blind to parity-violating signals. 
\paragraph*{Polarised Galactic foreground:} To break the degeneracy between $\alpha$ and $\beta$, one can exploit the fact that the Galactic foreground itself is polarised. The Galactic foreground emission is expected to be rotated only by $\alpha$, whereas the CMB is rotated by $\alpha+\beta$ \cite{Minami:2019ruj,Minami_2020,Minami:2020fin,Minami:2020odp,PhysRevLett.127.151301}. This method has reached a precision of $\sim 0.1^\circ$ \cite{Abitbol_2016,Diego_Palazuelos_2022}. However, this framework assumes primordial $EB=0$ and relies on modeling assumptions for Galactic foreground $EB$, which may in reality be intrinsically nonzero and sky-dependent, for example due to magnetically misaligned filamentary dust structures \cite{2021ApJ...919...53C}.
\paragraph*{Astrophysical polarised sources:} Bright, polarised point sources in the sky such as the Crab Nebula \cite{Aumont_2010,Kaufman_2016,Aumont_2020,Masi_2021} can be used as calibration standards. The current precision reached with this method does not exceed $0.5^\circ$ \cite{Takahashi_2010,Komatsu_2011}, due to uncertainties in measuring the intrinsic polarisation angle of the source itself.
\paragraph*{Instrumental calibration techniques:} Artificial calibrators such as wire-grids can measure the absolute polarisation orientation in situ, independently of sky assumptions \cite{Keating_2003,Takahashi_2010,Hinderks_2009,2012JLTP167936T,Murata_2023}. The anticipated precision for the SO SATs is $\sim 0.08^\circ$ \cite{Murata_2023}, with demonstrations on early data reported in \citet{Nakata_2026}. Moreover, \cite{Coppi:2025fmt} shows the promise of a drone-based polarisation calibrator.

\cref{summary} provides a non-exhaustive overview of current calibration techniques and of the assumptions on which they rely.
\begin{table*}[t]
\centering
\begin{tabular}{|c|c|c|c|l|}
\hline
Type of calibration & Precision & $C^{EB}_\ell=0$? & $\beta=0$? & References \\
\hline\hline

Self-calibration & $0.03^\circ$ & yes & yes &
\makecell[l]{
  \cite{Keating_2012}\\
  \cite{2016_parity}\\
  \cite{AtacamaCosmologyTelescope:2025blo}
} \\
\hline

Galactic foreground & $0.1^\circ$ & yes & no &
\makecell[l]{
  \cite{Minami:2019ruj}\\
  \cite{Minami_2020}\\
  \cite{Minami:2020fin}\\
  \cite{Minami:2020odp}\\
  \cite{PhysRevLett.127.151301}
} \\
\hline

\makecell{Astrophysical\\polarised sources}
& $0.3^\circ$ & no & no &
\makecell[l]{
  \cite{Aumont_2010}\\
  \cite{Komatsu_2011}\\
  \cite{Kaufman_2016}\\
  \cite{Aumont_2020}\\
  \cite{Masi_2021}\\
  \cite{Coppi:2025fmt}
} \\
\hline

\makecell{Artificial\\polarised sources}
& $0.5^\circ$ & no & no &
\makecell[l]{
  \cite{Ritacco:2023mgi}\\
  \cite{Coppi:2025fmt}\\
  \cite{BICEPKeck:2024cmk}\\
  \cite{polish2025improvedabsolutepolarizationcalibrator}\\
  \cite{2026FrASS..1371698C}
} \\
\hline

Wire grid * & $0.08^\circ$ & no & no &
\makecell[l]{
  \cite{Murata_2023}\\
  \cite{Nakata_2026}
} \\
\hline
\end{tabular}
\caption{Summary of existing polarisation angle calibration methods,
including their typical precision and the assumptions they require on
the parity-violating sky signal.
\newline
*\small{Note that the wire-grid calibration is currently planned only for the
SO SATs, and its achievable precision will be validated with additional
on-sky data.}}
\label{summary}
\end{table*}
In a nutshell, astrophysical and foreground-based methods are limited by modeling uncertainties; self-calibration is highly precise but assumption-dependent; and instrumental calibrators provide an external, assumption-light alternative. However, the latter are not available for most experiments, since they are particularly difficult to implement for space-based detectors and remain challenging even for ground-based instruments.

\section{Cross-correlation functions between different instruments}
\label{sec:method}
We now describe the cross-instrument approach, which complements these techniques by isolating the \emph{relative} misalignment $\alpha_i - \alpha_j$ between instruments.

\subsection{The method}
Consider two instruments labelled by $i,j$ observing the \emph{same} patch of sky.\footnote{No more information can be retrieved with three or more instruments, as shown in Appendix \ref{app:three_detectors}.} Our goal is to construct an estimator of the \emph{relative} misalignment angle by exploiting cross-correlations between independent instruments. This quantity is robust to the presence of parity-violating signals such as cosmic birefringence or primordial $C^{EB}_\ell$.\footnote{Note that in this analysis, any $EB$ correlations sourced by foregrounds are treated as part of the observed signal rather than separated from a possible cosmological component.} Importantly, the method cannot by itself disentangle a common absolute rotation shared by all instruments from a cosmological signal. If both an isotropic birefringence angle $\beta$ and a primordial $C^{EB}_\ell$ are allowed, cross-spectra alone cannot break this degeneracy. What the method \emph{can} do is precisely constrain the relative misalignment between two instruments. If one of them is independently and accurately calibrated, as is expected for the SO SATs \cite{Nakata_2026}, as well as for the BICEP instruments \cite{BICEPKeck:2024cmk,polish2025improvedabsolutepolarizationcalibrator}, this relative measurement can then be used to determine the misalignment angle of the second instrument.

Starting from the leakage equation \cref{leakage},  we can compute the correlation between different instruments $i\neq j$: $C_{\ell,\;ij}^{EE, o}$, $C_{\ell,\;ij}^{BB, o}$, and $C_{\ell,\;ij}^{EB, o}$ in terms of the primordial power spectra $C^{EE}_\ell$, $C^{EB}_\ell$ and $C^{BB}_\ell$  which are the same for all instruments as they are observing the same sky patch.\footnote{Power spectra are calculated under a common mask definition.} We find that, for $i\neq j$,
\begin{widetext}
\begin{equation}
\begin{cases}
      C^{EE,o}_{\ell,\;ij}= C^{EE}_\ell \cos(2\tilde{\alpha}_i) \cos(2\tilde{\alpha}_j)+  C^{BB}_\ell \sin(2\tilde{\alpha}_i) \sin(2\tilde{\alpha}_j)-C^{EB}_\ell \sin(2\tilde{\alpha}^+_{ij})\\ C^{BB,o}_{\ell,\;ij}= C^{EE}_\ell \sin(2\tilde{\alpha}_i) \sin(2\tilde{\alpha}_j)+  C^{BB}_\ell \cos(2\tilde{\alpha}_i) \cos(2\tilde{\alpha}_j)+C^{EB}_\ell \sin(2\tilde{\alpha}^+_{ij})\\C^{EB,o}_{\ell,\;ij}= C^{EE}_{\ell} \cos(2\tilde{\alpha}_i) \sin(2\tilde{\alpha}_j)-  C^{BB}_\ell \sin(2\tilde{\alpha}_i) \cos(2\tilde{\alpha}_j)+
      C^{EB}_\ell \cos(2\tilde{\alpha}^+_{ij})
    \end{cases},
    \label{correlation}
\end{equation}
\end{widetext}
where we define $\tilde{\alpha}^{\pm}_{ij} \equiv \tilde{\alpha}_{i} \pm \tilde{\alpha}_{j}$. Expressing $\tilde{\alpha}^{\pm}_{ij}$ in terms of the telescope misalignment $\alpha_i$ and birefringence angle $\beta$, as defined in \cref{alpha_total}, 
\begin{equation}
\begin{cases}
\tilde{\alpha}^+_{ij}=\alpha_i+\alpha_j+2\beta \\
    \tilde{\alpha}^-_{ij}=\alpha_i-\alpha_j =  \alpha^-_{ij}\;,
\end{cases}
\label{delta_ij}
\end{equation}
shows that  $\tilde{\alpha}^-_{ij}$ does not depend on $\beta$ \cite{BICEPKeck:2024cmk}. This will be crucial for the purpose of the proposed method, and from now on we shall use interchangeably $\alpha^-_{ij}$ and $\tilde{\alpha}^-_{ij}$ since they are equal. 

For each pair of instruments $i \neq j$, the following cross-correlations can be constructed from the observed data:
\begin{equation}
     \begin{cases}
        C^{EE,o}_{\ell,\;ij}+C^{BB,o}_{\ell,\;ij}=(C^{EE}_{\ell}+C^{BB}_{\ell}) \cos\left(2\alpha^-_{ij}\right)\\
        C^{EB,o}_{\ell,\;ji}-C^{EB,o}_{\ell,\;ij}= (C^{EE}_{\ell}+C^{BB}_{\ell}) \sin \left( 2\alpha^-_{ij} \right)\;.
    \end{cases}
    \label{totaleq_1}
\end{equation}

These relations now allow a complete determination of $\alpha^-_{ij}$. Although $C^{EE}_{\ell}$ and $C^{BB}_{\ell}$ are unknown, they correspond to the same sky signal and are therefore identical for all detectors \footnote{At this stage, we do not exploit cross-frequency correlations and therefore assume that $C_\ell^{EE}$, $C_\ell^{BB}$, and $C_\ell^{EB}$ are identical for the two instruments. This is a good approximation for the SAT–LAT comparison, since the instruments share the same nominal frequency bands. The SO and Planck frequency channels, however, differ by a few GHz. Around the relatively foreground-clean CMB window at $143/145 \,\mathrm{GHz}$, the resulting differences are expected to be small, and we therefore expect our method to remain applicable in this regime.}. One obtains
\begin{equation}
 \alpha^-_{ij}=\underbrace{\frac{1}{2}\arctan\!\left(\dfrac{C^{EB,o}_{\ell,\;ji}-C^{EB,o}_{\ell,\;ij}}{C^{EE,o}_{\ell,\;ij}+C^{BB,o}_{\ell,\;ij}}\right)}_{\equiv R_{ij}} \;.
    \label{delta_ij_minus}
\end{equation}

The ratio $R_{ij}$ provides a direct estimate of the difference between the telescope misalignment angles using only their cross-spectra, without requiring prior assumptions about the intrinsic $C^{EB}_\ell$ spectrum or the presence of cosmic birefringence $\beta$. This makes the estimator particularly powerful: it targets precisely the observable that is both measurable and free from degeneracy with parity-violating physics. It therefore provides a probe of the \emph{relative} polarisation-angle calibration that is independent of assumptions on the intrinsic $EB$ spectrum and birefringence. Note that this method is straightforward, free to implement and can be applied to any pair of instruments observing the same patch of sky.

\subsection{Calibration estimation}
In this section, we use analytic derivation and simulated data to illustrate the estimator's performance and its expected precision under the forecast assumptions. The forecasts assume perfectly characterised beams and fully beam-deconvolved spectra. This represents an idealised scenario, and a detailed assessment of the impact of beam uncertainties is left for future work.
\subsubsection{Analytical evaluation}
Let us now consider the case where the first telescope is one of the SO SATs, with a misalignment angle $\alpha_1$; and the second is either the SO LAT or \textit{Planck}, with a misalignment angle $\alpha_2$. From \cref{delta_ij_minus}, the relative angle $\alpha_2-\alpha_1$ can be determined directly from the data, using the cross-correlation between experiments. From \cref{delta_ij_minus}, we find that the standard deviation on $\alpha_2$ is
\begin{equation}
    \sigma^2_{\alpha_2} = (\sigma_{\alpha_1})^2+(\sigma_{R_{12}})^2
\end{equation}
with $\sigma_{\alpha_1}$ the uncertainty on the calibration of the SAT, and $\sigma_{R_{12}}$ the uncertainty due to the pixel noise of different instruments. The best mechanical calibration method being implemented for the SO SATs is expected to have a systematic error of $\sigma_{\alpha_1}=0.08^\circ$  \cite{Murata_2023,Nakata_2026}, so this is what we will assume from now on. For $\sigma_{R_{12}}$, we can propagate the error of \cref{delta_ij_minus} in the small angle limit: 
\begin{equation}
    \frac{\sigma^2_{\alpha^-_{ij}}(\ell)}{\left( \alpha_{ij}^-\right)^2}= \dfrac{
    \Xi_{\ell,ij}^{EB,o}+\Xi_{\ell,ji}^{EB,o}}{(C^{EB,o}_{\ell, \;ij}-C^{EB,o}_{\ell, \;ji})^2}+\dfrac{\Xi_{\ell,ij}^{EE,o}+\Xi_{\ell,ij}^{BB,o}}{(C^{EE,o}_{\ell, \;ij}+C^{BB,o}_{\ell, \;ij})^2} \;,
    \label{sigma_alpha_ij}
\end{equation}
where $\Xi$ denotes the variances of the power spectra. These can be estimated 
using the Knox approximation to compute the Gaussian covariance 
\cite{Knox:1995dq,Bowden_2004}:
\begin{equation}
\begin{aligned}
\Xi^{XY,X'Y'}_\ell
= & \ \frac{1}{(2\ell+1) f_{\rm sky}} \\
\times & \Big[
\left( C^{X Y^{\prime}}_\ell +N^{X Y^{\prime}}_\ell \right)
\left(C^{Y X^{\prime}}_\ell+N^{Y X^{\prime}}_\ell \right) \\
& +
\left(C^{X X^{\prime}}_\ell+N^{X X^{\prime}}_\ell \right)
\left(C^{Y Y^{\prime}}_\ell+N^{Y Y^{\prime}}_\ell \right)
\Big],
\end{aligned}
\label{eq:knox_general}
\end{equation}
where $X,Y,X',Y'\in\{T,E,B\}$, $C^{XY}_\ell$ denotes the signal power spectrum, $N^{XY}_\ell$ the corresponding noise contribution and $f_\text{sky}$ the sky coverage.
For cross-spectra between two instruments $i$ and $j$, assuming uncorrelated noise (i.e. $N^{XY}_{\ell, \;ij}=0$),
the diagonal covariance for the measured cross-spectrum $C^{XY,o}_{\ell,\; ij}$
reduces to 
\begin{equation}
\begin{aligned}
\Xi_{\ell, \; ij}^{XY,o}
&= \ \frac{1}{(2\ell+1) f_{\rm sky}} \\
\times  \Big[ (C^{XY,o}_{\ell,\;ij})^2
 + &(C^{XX,o}_{\ell,\;ii}+N^{XX}_{\ell,\;i})
(C^{YY,o}_{\ell,\;jj}+N^{YY}_{\ell,\;j})\Big],
\end{aligned}
\label{eq:knox_cross_uncorr}
\end{equation}
where $i,j$ label the telescope and $f_\text{sky}$ is the largest overlapping sky portion between them. While the assumption $N^{XY}_{\ell,ij}=0$ is reasonable for genuinely independent experiments, partial noise correlations may arise in the SAT--LAT comparison. Since the instruments are located at the same site, they observe through partially shared atmospheric conditions and may also employ overlapping scan strategies. Assessing the resulting correlations would require map-level simulations that capture their joint observing conditions, or data-driven estimates from observations. Such an analysis is beyond the scope of this study and is left for future work \cite{lonappancoming}. Among the instrumental combinations considered here, the SAT--LAT forecast is therefore the most susceptible to correlated-noise effects.

\begin{figure}[]
    \centering
    \includegraphics[width=1\linewidth]{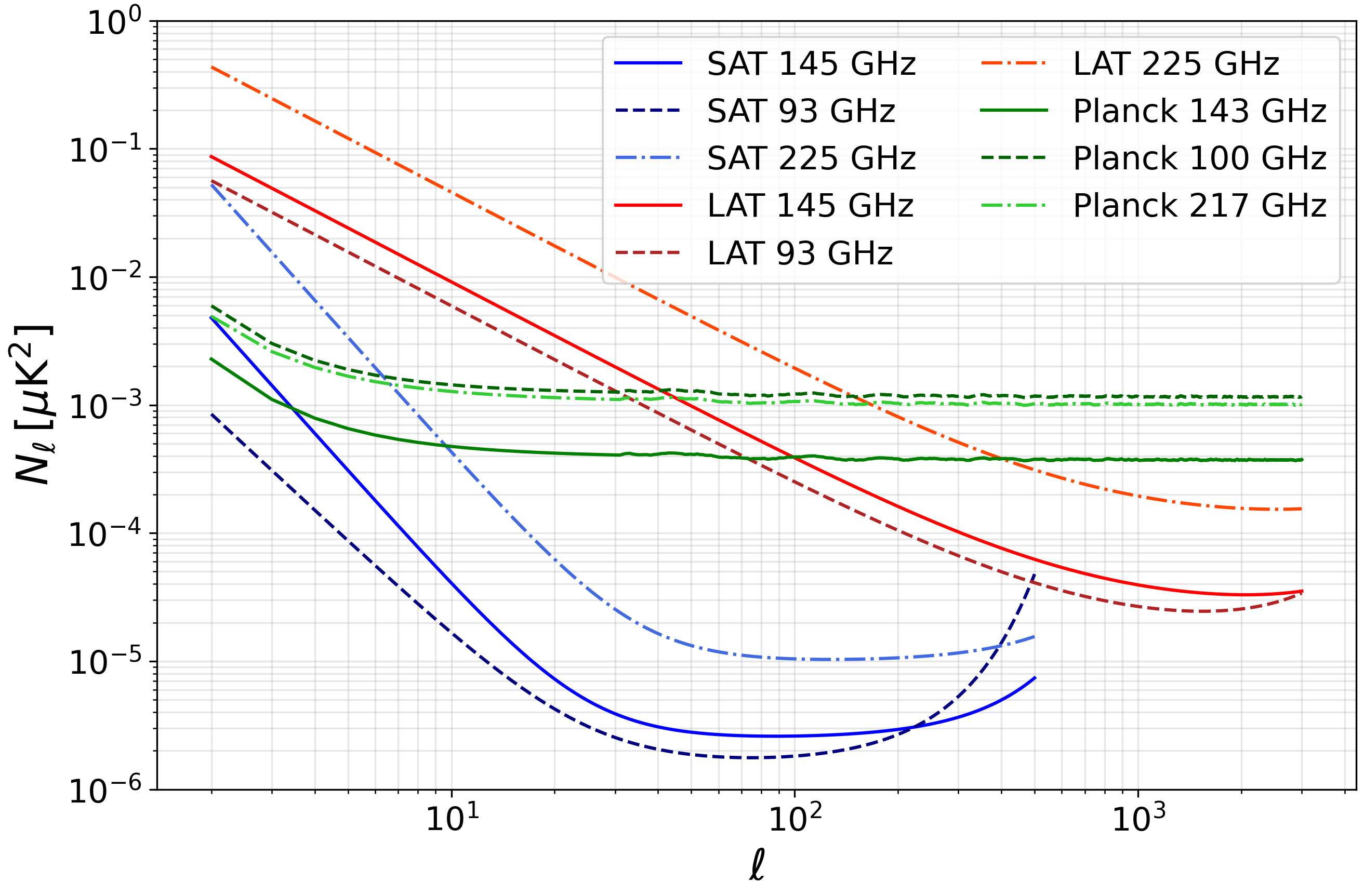}
    \caption{Forecast noise power spectra for the SO SATs and LAT, as well as achieved noise for \textit{Planck}. 
    We restrict to the SAT analysis multipole range $\ell \in[30,500]$, which captures the large angular scales targeted by SAT observations and overlaps at the high-$\ell$ end with the LAT targeted scales \cite{Ade_2019}. Beam effects are included for both SO SATs and LAT.}
\label{noise_comparison}
\end{figure}

We assume that the limiting sky coverage is always due to the SO SATs \cite{Ade_2019, SimonsObservatory:2025avm}, 
so we use $f_\text{sky}=0.1$ from now on. We generate the expected noise curves for SO using the noise model described in \cite{Ade_2019}.\footnote{We use the code on Github \href{https://github.com/simonsobs/so_noise_models/tree/master}{\textcolor{black}{\faGithub}}, at the url \url{https://github.com/simonsobs/so_noise_models/tree/master}.} To model the \textit{Planck} noise, we use the pixel covariance data from the \textit{Planck} legacy website. \footnote{The code to process \textit{Planck} products is available on Github \href{https://github.com/crigouzzo/CMB_Polarisation_Calibration}{\textcolor{black}{\faGithub}}, at the url \url{https://github.com/crigouzzo/CMB_Polarisation_Calibration}} A comparison of the forecast noise levels for the SO SATs and LAT, and those achieved for \textit{Planck}, is shown in \cref{noise_comparison}.

Finally, if we assume the $\ell$ mode to be independent, we can use the inverse-variance weighted average \cite{Mandel1964-MANTSA} to estimate the \emph{total} error on $R_{ij}$:
\begin{equation}
    \sigma^2_{R_{ij}}=\left(\sum_\ell\frac{1}{\sigma_{\alpha^-_{ij}}^2(\ell)}\right)^{-1}\;.
    \label{inverse_variance}
\end{equation}
With all ingredients in place, we proceed to compute the expected uncertainty on $R_{12}$, taking an SO SAT as the reference instrument and either the LAT or \textit{Planck} as the second dataset. For the $143$\,GHz channel, we find:
\begin{equation}
\begin{cases}
    \sigma_{R_{12}} \simeq 0.07 ^ \circ & \text{for SAT with LAT} \\
    \sigma_{R_{12}} \simeq 0.15 ^ \circ & \text{for SAT with \textit{Planck}}
\end{cases}
\label{standard_deviation}
\end{equation}
which implies that the LAT/ \textit{Planck} can be calibrated to a precision of: 
\begin{equation}
\begin{cases}
\sigma_{\alpha_2} \simeq \sqrt{(0.08 ^ \circ)^2+(0.07^ \circ)^2} \simeq 0.10 ^ \circ & \text{for LAT}\\
\sigma_{\alpha_2} \simeq \sqrt{(0.08 ^ \circ)^2+(0.15^ \circ)^2} \simeq 0.17  ^ \circ & \text{for \textit{Planck}}
\end{cases}    
\label{propagation_uncertainties}
\end{equation}
We can also compare the results in different frequency channels, as summarized in \cref{freq_calibration}. Clearly, our method achieves higher precision when applied to the LAT, thanks to the lower noise power spectra, see \cref{noise_comparison}.
\begin{table}[h]
\hspace{0.1\linewidth}
\begin{tabular}{|c|c|c|c|}
\hline
$f_\text{SO}$ [GHz] & $f_\text{\textit{Planck}}$ [GHz] & \textit{Planck} [$^\circ$]& LAT [$^\circ$] \\ \hline
$93$      & $100$     & 0.28       &  0.10  \\ \hline
$145$     & $143$     & 0.17      &  0.10   \\ \hline
$225$     & $217$     &  0.27    &  0.16   \\ \hline
\end{tabular}
\caption{Comparison of the expected calibration precision $\sigma_{\alpha_2}$ for different frequencies, for both SAT \& \textit{Planck} and SAT \& LAT. This assumes a SAT calibration uncertainty of 0.08$^{\circ}$.}
\label{freq_calibration}
\end{table} 
\subsubsection{Simulation evaluation}
Let us now test the approach against simulations. 
\paragraph*{Procedure:} We realise a total of $5000$ simulations, with modes $\ell \in [30,500]$. First, we choose fiducial values for the total polarisation misalignment angle $\tilde{\alpha}=\alpha+\beta$, which combines telescope miscalibration $\alpha$ and cosmic birefringence $\beta$. Telescope calibration is expected to be within $0.1^\circ$, and current indications suggest birefringence may be of a similar order \cite{Diego-Palazuelos:2025dmh,Diego_Palazuelos_2022}. We adopt $\tilde{\alpha}_1=0.057^\circ$ for SAT and $\tilde{\alpha}_2=-0.069^\circ$ for \textit{Planck}/LAT, and verify that the method remains valid for angles up to $\pm 5^\circ$.

We generate fiducial sky power spectra $C_\ell^{EE}$ and $C_\ell^{BB}$, assuming negligible $C_\ell^{EB}$, and compute the corresponding \emph{observed} spectra using \cref{correlation}. Instrumental noise power spectra are then added for SAT, LAT, and \textit{Planck} in matching frequency bins, using publicly available noise models. The covariance of the spectra between instruments is computed using the Knox formula (\cref{eq:knox_general}).

For each multipole $\ell$, Gaussian noise realisations are drawn according to this covariance and added to the rotated spectra. For each realisation we evaluate the estimator for the relative misalignment $\alpha_1-\alpha_2$ given in \cref{delta_ij_minus}. Repeating this procedure over many realisations allows us to determine both the standard error of the estimator and its multipole-dependent variance. Assuming independent multipoles, the final uncertainty is obtained through inverse-variance weighting \cref{inverse_variance}. The resulting uncertainties are summarized in the Appendix in \cref{tab:comparison_simulation}.

\begin{figure}
    \centering
    \includegraphics[width=0.7\linewidth]{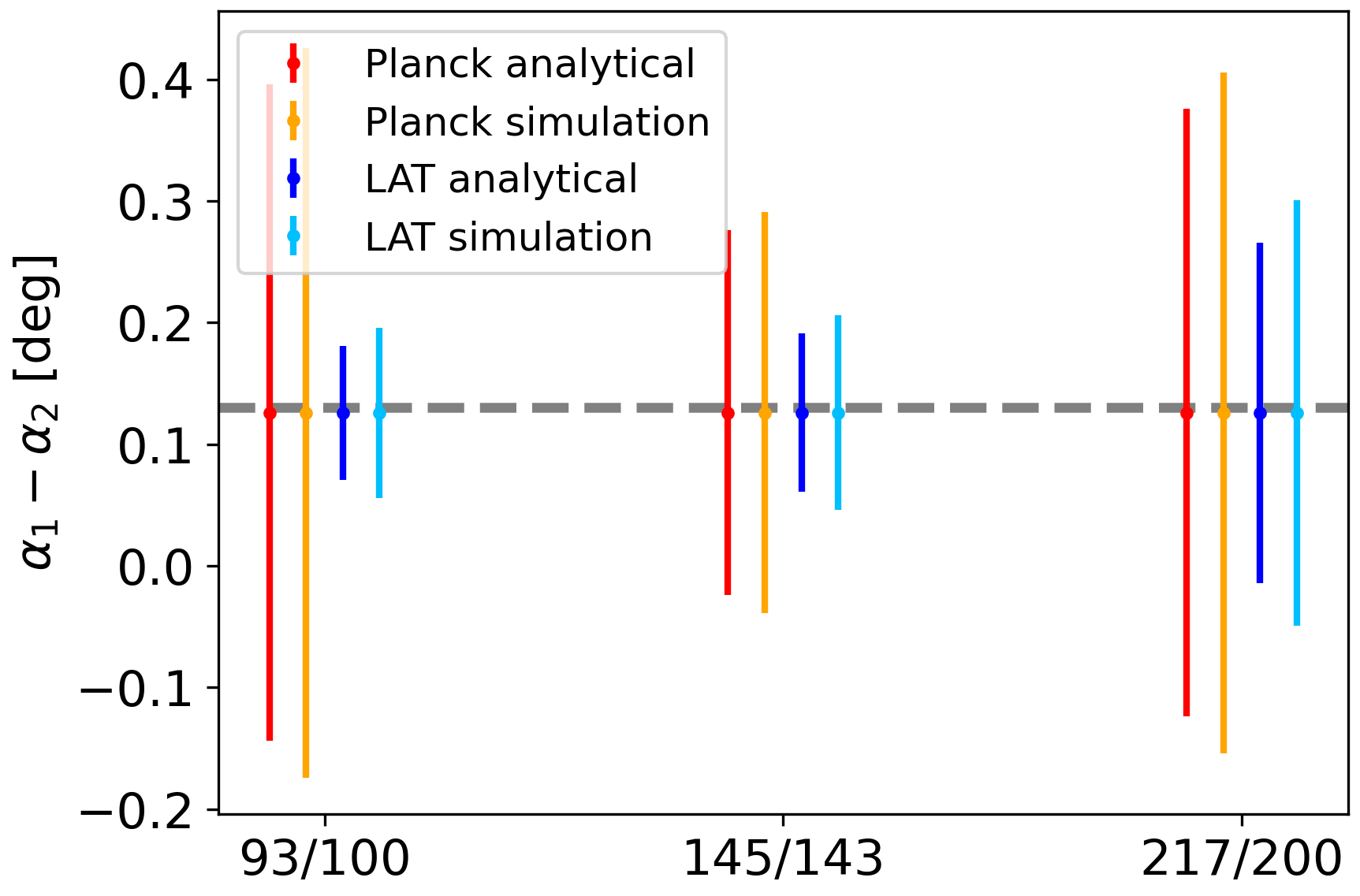}
    \caption{Comparison of the calibration precision $\sigma_{R_{ij}}$ between simulation and analytical result for different frequencies, for both SAT \& \textit{Planck} and SAT \& LAT. The frequency channels are $93$,$145$ and $225$ GHz for the Simons observatory and  $100$, $143$ and $217$ GHz for \textit{Planck}. For the simulations, we chose $\tilde{\alpha}_1=0.057^\circ$ and $\tilde{\alpha}_2=-0.069^\circ$, and show in a grey dashed line the corresponding fiducial value $\alpha_1-\alpha_2=0.13^\circ$. Values for the standard deviation can be found in the Appendix in \cref{tab:comparison_simulation}.}
    \label{fig:comparison_simulation}
\end{figure}

\paragraph*{Results}We find that the estimator suggested in \cref{delta_ij_minus} accurately recovers the difference between telescope misalignment angles, with a standard error 
of $0.009^\circ$. Unbiased-ness is further verified by testing the estimator across a range of injected angle values, from $-5^\circ$ up to $5^\circ$. Results for the standard deviation are shown in \cref{fig:comparison_simulation}, with a comparison with the analytical method. It shows good agreement between 
simulations and the theoretical expectation, with discrepancies of order 
$0.05^\circ$ depending on frequency and instrument.

\paragraph*{Potential sources of leakage} Beam mismatch, ellipticity and masking can induce various forms of leakage between the \(E\) and \(B\) modes. For Planck, these effects are relatively well controlled, and the beam can be safely approximated as Gaussian over the range \(\ell<1500\) relevant to our analysis \cite{Planck:2013dwd,2016A&A594A7P,Naess:2020wgi}. For the SO SATs and LAT, the impact of beam systematics is still being characterised using complementary methods; see, for example, \cite{Dachlythra_2024,Chesmore:22}. Instrument-dependent polarized beam asymmetries or differential beam mismatch can induce \(E\)-to-\(B\) leakage that does not cancel in the antisymmetric combination and may mimic a relative, potentially scale-dependent, polarisation rotation. Quantifying this effect requires instrument-specific beam models, scan strategies, and map-level simulations, and will be necessary for an application to real data. An $\ell$-dependent recovered angle would provide a useful diagnostic of such contamination.
\section{Constraining Parity-Violation}
\label{sec:parity}
Having determined the telescope misalignment in \textit{Planck}/LAT, we can now pursue two approaches to assess whether parity-violating physics is present. We assume that the effect of isotropic cosmic birefringence and a primordial $C^{EB}_\ell$ power spectrum can be separated, which to a good approximation is the case when there is a large asymmetry between $C^{EE}_{\ell}$ and $C^{BB}_{\ell}$, see \cref{E_o_B_o} \cite{Komatsu_2022}. This means that we can separately investigate the case where $\beta=0$, to constrain the primordial $C^{EB}_\ell$; and the case where $C^{EB}_\ell=0$ to give an estimate of $\beta$. We will start with the latter.
\subsection{Constraining isotropic birefringence}
\begin{figure}
    \centering
\includegraphics[width=1\linewidth]{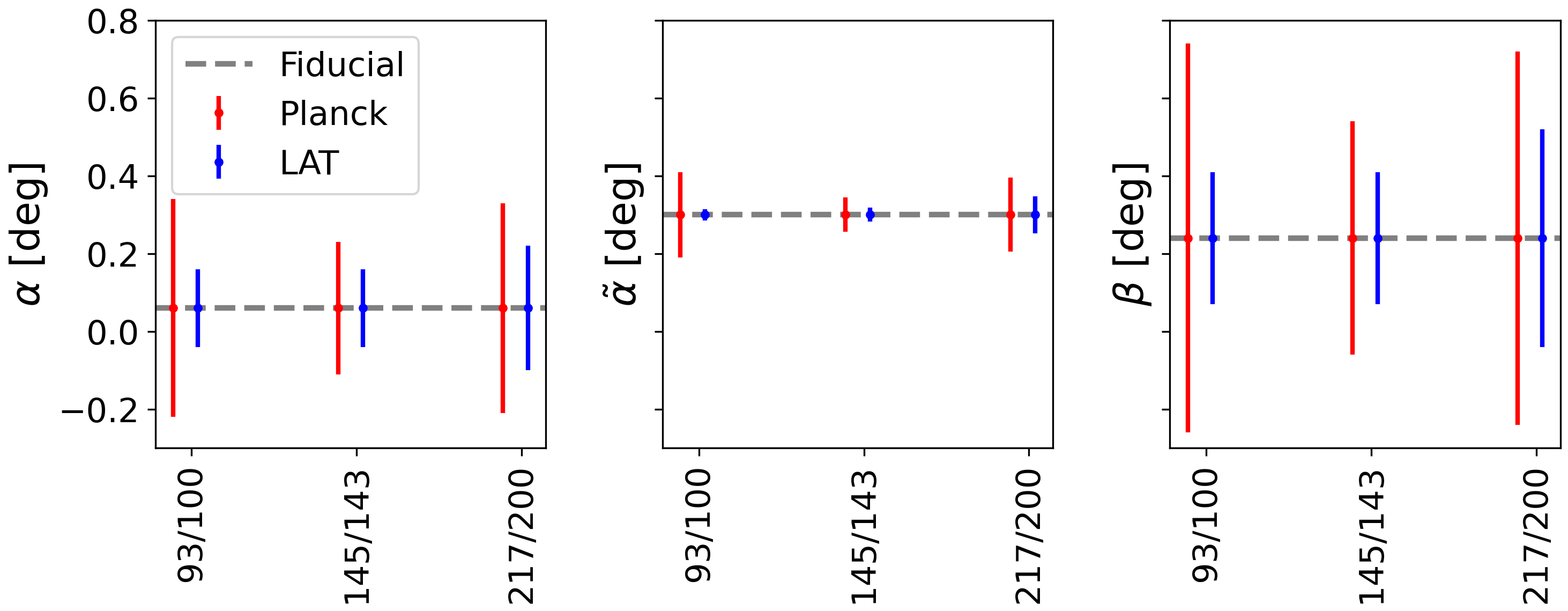}
    \caption{Comparison of the standard deviation for the telescope angle $\alpha$, the total angle $\tilde{\alpha}$ and the cosmic birefringence angle $\beta$ at different frequencies, for both \textit{Planck} and LAT. The frequency channels are $93$,$145$ and $225$ GHz for the Simons observatory and  $100$, $143$ and $217$ GHz for \textit{Planck}. Values for the standard deviation can be found in \cref{birefringence_LAT_Planck}.}
    \label{birefringence_total}
\end{figure}

Assuming the primordial $C_\ell^{EB}=0$, we can solve for the total misalignment angle within each instrument: 
\begin{equation}
4\tilde{\alpha}_i=\arctan \left(\frac{2 C^{EB,o}_{\ell,ii}}{C^{EE,o}_{\ell,ii}-C^{BB,o}_{\ell,ii}}\right), 
\end{equation}
where $\tilde{\alpha}_i=\alpha_i+\beta$  includes both the instrumental and birefringence misalignment, and this time we only consider the power spectrum within one instrument ($i=j$): $C^{XY,o}_{\ell, \; ii}$. We can evaluate the standard deviation of $\tilde{\alpha}_i$:
\begin{equation}
\begin{split}
    4  \sigma^2_{\tilde{\alpha}_i}(\ell) & \simeq \frac{ \Xi_{\ell, i i}^{EB,o}}{ (C^{EE,o}_{\ell, ii}- C^{BB,o}_{\ell, ii})^2} 
    \quad \\ & +\left( \frac{ C^{EB,o}_{\ell, ii} }{C^{EE,o}_{\ell, ii}-C^{BB,o}_{\ell, ii}}\right)^2
    \left( \Xi_{\ell, i i}^{EE,o}+\Xi_{\ell, i i}^{BB,o}\right) \;.
    \end{split}
    \end{equation}
Then we can estimate the uncertainty on the isotropic cosmic birefringence through $\sigma^2_\beta \simeq 3 \sigma^2_{\alpha_i}+ \sigma^2_{\tilde{\alpha}_i}$ \footnote{We used that for any two random variables, $\operatorname{Var}(X+Y)=\operatorname{Var}(X)+\operatorname{Var}(Y)+2 \operatorname{Cov}(X, Y)$. In the case of $\alpha_i$ and $\tilde{\alpha}$, we assume that $\operatorname{Cov}(\alpha_i+\beta, \alpha_i)\simeq \operatorname{Var}(\alpha_i)$ since the telescope misalignement $\alpha_i$ is independent of $\beta$.}. Let us simulate an example where the telescope misalignment for both LAT and \textit{Planck} is $\alpha_i=0.06^\circ$, and $\beta=0.24^\circ$, following indications in \cite{Diego_Palazuelos_2022,Diego-Palazuelos:2025dmh}. Then, the standard deviation for the telescope angle $\alpha_i$, the total angle $\tilde{\alpha}_i$ and the cosmic birefringence angle $\beta$ is shown in \cref{birefringence_total}, where we dropped the $i$ subscript for clarity.
 \cref{birefringence_total} shows that the LAT achieves smaller statistical uncertainties on $\beta$ than \textit{Planck}, primarily due to its lower polarisation noise over the overlapping sky region. Furthermore, \cref{birefringence_total} indicates that the dominant systematic uncertainty arises from the telescope misalignment angle $\alpha$, which therefore sets the limiting threshold on how well the cosmic birefringence parameter $\beta$ can be constrained. 
\subsection{Constraining primordial EB}
Next, we can assume that $\beta=0$, and constrain the primordial $C^{EB}_\ell$ power spectrum. In this case, $\tilde{\alpha}_i=\alpha_i$. Using only one instrument, from \cref{E_o_B_o}, we can read the primordial $C^{EB}_\ell$ power spectrum \cite{Minami:2019ruj}:
 \begin{equation}
     C^{EB}_\ell \simeq C^{EB,o}_{\ell,ii} - 2\alpha_i \left(C^{EE,o}_{\ell,ii}-C^{BB,o}_{\ell,ii} \right).
 \end{equation}
 Then, the error propagation for the intrinsic $EB$ becomes: 
 \begin{equation}
 \begin{split}
      \Xi^{EB}_{\ell} & \simeq \Xi_{\ell, i i}^{EB,o}+4\alpha_i^2 \left( \Xi_{\ell, i i}^{EE,o}+\Xi_{\ell, i i}^{BB,o}\right) \\ &+4\left(C^{EE,o}_{\ell,ii}-C^{BB,o}_{\ell,ii} \right)^2\sigma^2_{\alpha_i} \;.
     \end{split}
     \label{intrinsic_EB}
 \end{equation}
 The first and second term encode the error due to the measurement of the $EE$, $BB$ and $EB$ power spectrum, through the Knox formulae. In particular, the second term is the total error on $EE$ and $BB$ rotated by the telescope misalignment. The third term is due to the error on the calibration, which we computed previously, for example in \cref{propagation_uncertainties}.

\begin{figure}[]
    \centering
    \includegraphics[width=1\linewidth]{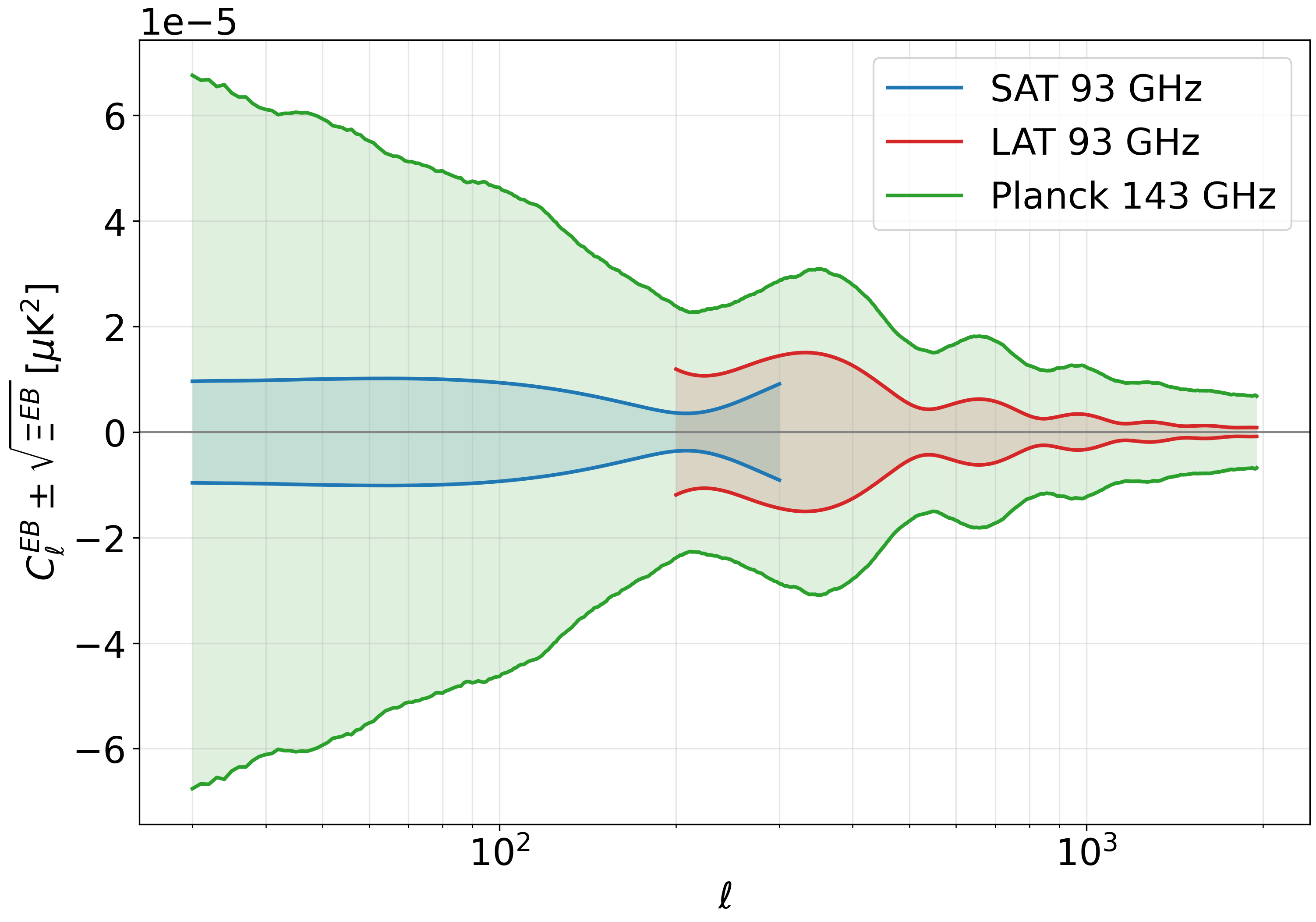}
\caption{Standard deviation on the primordial $C^{EB}_\ell$ power spectrum for \textit{Planck}, Simons Observatory SATs and LAT for different frequency channels. We used the frequency channel where the noises are the lowest: $93$ GHz for LAT and SATs, and $143$ GHz for \textit{Planck}.}
    \label{error_EB}
\end{figure}

 We can now simulate a vanishing $C^{EB}_\ell$ and plot the associated standard deviation given by \cref{intrinsic_EB}. The results are shown in \cref{error_EB} for the three different instruments.\footnote{We chose to plot frequency channels that yields the lowest standard deviation for $\alpha_i$ to show the best constraints.} This gives us an idea of the amplitude and shape necessary to be detectable with our current calibration method. 
From \cref{error_EB} we observe that the SAT is achieving the best precision, but has a very limited $\ell$ range. On the other hand, \textit{Planck} and the LAT can provide information about the $EB$ power spectrum at a higher $\ell$. Interestingly, the uncertainties on the primordial $C^{EB}_\ell$ signal for \textit{Planck} and the LAT are comparable, despite \textit{Planck}'s substantially poorer calibration precision. This is primarily due to \textit{Planck}'s broader sky coverage, which compensates for its larger calibration uncertainty by suppressing the overall covariance. These results highlight the importance of improved post-calibration for \textit{Planck} in enhancing its scientific return. Finally, let us note that the largest source of uncertainty comes from the pixel covariance term: $\Xi_{\ell, i i}^{EB,o}$, which indicates that the calibration uncertainty has reached sharp enough precision.

\section{Discussion and outlook}
\label{sec:discussion}
Let us now outline the strengths and weaknesses of this approach. The central goal was to describe a calibration technique that does not require assuming vanishing $\beta$ and $C^{EB}_\ell$. In \cref{summary2}, we compare it with existing methods that similarly avoid these assumptions. A key requirement of this approach is the availability of at least one well-calibrated instrument. In this work, we assume that the Simons Observatory SATs achieve a calibration precision of $\sim0.08^\circ$ using a wire-grid system. 
However, this level of performance has not been fully demonstrated on-sky, and should therefore be regarded as an assumption in our forecasts.
\footnote{We have carried out the analysis for \textit{Planck} and the LAT only, however the approach could be used for all other CMB polarisation experiments that share a sufficient sky overlap with SO SATS, such as the Atacama Cosmology Telescope \cite{AtacamaCosmologyTelescope:2025nti}, LiteBIRD \cite{LiteBIRD:2025yfb}, WMAP \cite{WMAP:2012fli}, BICEP \cite{2010SPIE.7741E..1GO} etc.}.
If the wire grid calibration for the SATs were to underperform for any reason, the precision of our method would worsen, as indicated by \cref{propagation_uncertainties}. Encouragingly, multiple calibration strategies are being pursued for the SATs, including drone-based systems, which offer a promising route to achieving high-precision, on-sky polarisation calibration. In addition, our approach requires that the telescopes share some overlap in sky coverage, which is not always guaranteed.
\begin{table}[]
\hspace{0.15\linewidth}
\begin{tabular}{|c|c|}
\hline
Type of calibration & Precision  [$^\circ$] \\ \hline
Wire grid  (SO SAT)      & $0.08$        \\ \hline
Polarised sources   & $0.50$        \\ \hline
Correlation SATs + LAT * & $0.10$          \\ \hline
Correlation SATs + \textit{Planck} * & $0.17$         \\ \hline
\end{tabular}
\caption{Summary of calibration methods that do not rely on model assumptions. The techniques introduced in this work are marked with an asterisk. For these forecasts, we only use the $143/145 \,\mathrm{GHz}$ channel.}
\label{summary2}
\end{table}

On the other hand, a major advantage of this calibration method is that its effectiveness will naturally improve over time. As more instruments with precise polarisation calibration come online, this method enables retrospective cross-calibration of existing datasets, enhancing their scientific value at minimal additional cost. This highlights the potential value of developing a dedicated ``\emph{super-calibrator}'': an instrument that does not need to survey the entire sky but is instead optimised for achieving exceptionally low calibration errors, which can then be used to recalibrate other CMB polarisation experiments. This illustrates a possible synergy between ground-based and space-based telescopes---combining a well-controlled, specialised calibration instrument with a larger-scale observational mission \cite{2026FrASS..1371698C}.

This paper demonstrates the basics of this calibration method; however, there remains significant scope for improving the error estimation. One potential avenue is to utilize all available frequency channels and perform a combined analysis. In the case of the polarised Galactic foreground calibration method, for instance, incorporating all frequency channels led to a twofold reduction in the error, see \cite{Minami:2019ruj} in comparison with \cite{Minami:2020fin}. A more detailed implementation of this SAT–LAT cross-calibration approach, including map-based simulations, an exploration of different calibration regimes, updated SO noise models, foreground treatment, will be presented in an upcoming Simons Observatory paper \cite{lonappancoming}. 
\section*{Conclusion}
In this work, we have examined the impact of polarisation-angle miscalibration as a key limitation in extracting parity-violating signals from the CMB. We described a method that uses cross-correlations between experiments to determine their relative polarisation-angle misalignment, $\alpha^-_{ij}$, without assuming vanishing isotropic cosmic birefringence or primordial $EB$ correlations. By leveraging cross-correlations between different experiments \cite{BICEPKeck:2024cmk}, we demonstrated that any datasets that share some sky coverage can be recalibrated using the SO SATs.

The central result is that an antisymmetric combination of cross-instrument $EB$ spectra isolates this relative angle while remaining insensitive
to both birefringence and intrinsic $EB$ signals, as shown in \cref{delta_ij_minus}. This identifies the specific observable that can be robustly extracted from two experiments at the level of two-point functions, and provides a direct estimator for it. While cross-correlations are widely used in CMB analyses, this formulation makes explicit how they can be used to calibrate polarisation angles without imposing assumptions that would otherwise remove the very signals of interest.

Our forecasts show that, assuming a well-calibrated reference instrument such as the Simons Observatory SATs, this method can achieve polarisation-angle uncertainties of $0.10^\circ$ for the LAT and $0.17^\circ$ for \textit{Planck} at 145/143 GHz (including both systematic and statistical errors). This demonstrates that cross-instrument calibration can significantly improve the effective calibration of existing datasets, thereby enhancing their sensitivity to parity-violating physics. A key requirement of this approach is the availability of at least one accurately calibrated instrument to anchor the relative measurements. In this work, we have assumed that the SATs achieve a calibration precision of $\sim 0.08 ^\circ$, which is the predicted precision using a wire-grid system. However, this level of performance has not yet been demonstrated with on-sky data and should therefore be regarded as an assumption in our forecasts. Encouragingly, additional calibration strategies, including drone-based systems, are being developed and offer promising avenues for achieving high-precision absolute calibration.

This approach enables a significant reduction of systematic uncertainties and enhances our ability to constrain or detect signatures of parity-violating physics in the early universe. More generally, this method highlights the potential of cross-instrument analyses to improve polarisation calibration across experiments. As future surveys with increasing sensitivity and improved calibration strategies become available, such approaches will play an important role in maximizing the scientific return of CMB polarisation measurements, particularly in the search for parity-violating physics beyond the standard cosmological model.

\paragraph*{Acknowledgments:} We are very grateful to Kam Arnold, Carlos Herv\'ias-Caimapo, Jo Dunkley, Lam Hui, Mudit Jain, Anto Lonappan and David Marsh for discussions and insightful feedback. We would also like to thank the organisers of the GC2024 conference at the Yukawa Institute where this work was originally conceived in a discussion between the authors and Jo Dunkley. C.R. acknowledges support from the
Science and Technology Facilities Council (STFC), as well as support from the Cromwell fellowship.

\appendix
\crefalias{section}{appendix}

\section{Isotropic cosmic birefringence production}
\label{iso_birefringence}
We demonstrate how the presence of parity-violating interactions affecting photons can be probed through the rotation of their polarisation plane. Consider a parity violating pseudoscalar field $\phi$, that couples to photons through a Chern-Simons term $\phi F_{\mu \nu} \tilde{F}^{\mu \nu}$, where $\tilde{F}_{\mu \nu}= 1/2 \epsilon_{\mu \nu \alpha \beta } F^{\alpha \beta}$ \cite{Harari:1992ea,Komatsu_2022}: 
\begin{equation}
\begin{split}
\mathcal{L} &=-\frac{1}{4} F_{\mu \nu} F^{\mu \nu}+\frac{1}{2} \partial_\mu \phi \partial^\mu \phi+\frac{1}{4} g \phi F_{\mu \nu} \tilde{F}^{\mu \nu} \\ &= \frac{1}{2}\left[\left(\boldsymbol{E}+\frac{g}{2 } \phi \boldsymbol{B}\right)^2-\left(\boldsymbol{B}-\frac{g}{2 } \phi \boldsymbol{E}\right)^2\right]+\mathscr{O}\left(g^2\right) \, .
\end{split}
\label{CS_term}
\end{equation}
\cref{CS_term} proves that it is now the combinations $\boldsymbol{E}+\frac{g}{2 } \phi \boldsymbol{B}$ and $\boldsymbol{B}-\frac{g}{2 } \phi \boldsymbol{E}$ that propagates freely and oscillates along a constant direction, not simply $\boldsymbol{E}$ and $\boldsymbol{B}$. In the simple case of a spatially homogeneous field $\phi(t)$, the plane of linear polarisation will be rotated by an angle 
\begin{equation}
    \beta =(g / 2)\left[\phi\left(t_0\right)-\phi\left(t_{\mathrm{LS}}\right)\right],
    \label{def_beta}
\end{equation}
 where $\phi(t_0)$ and $\phi\left(t_{\mathrm{LS}}\right)$ are the values of the field nowadays and at the surface of last scattering \cite{Komatsu_2022}. We showed how such a term would affect right- and left-handed photons asymmetrically, leading to a rotation of the polarisation plane as photons propagate from the last scattering surface to the present.

\section{What we \textit{cannot} do with cross correlating between different detectors:} \label{app:three_detectors}
One may think that we could directly constrain the telescope misalignment angles $\alpha_i$ from cross correlating the $E$- and $B$-modes of three or more instruments. Maybe we could compute $\tilde{\alpha}_{ij}^+$, and then deduce the misalignment angle $\alpha_i$? 
Unfortunately, this is not possible as we will now demonstrate.

Let the observed polarisation for the $i$-th instrument and true polarisation  be ${\bf u}_i = (E^o_i ,\; B^o_i )^\text{T}$ and  ${\bf v} =(E ,\; B )^\text{T}$ respectively. As shown in \cref{leakage}:
\begin{equation}
    {\bf u}_i = R(2\tilde{\alpha}_i){\bf v}\label{eqn:uRv} \;.
\end{equation}
 We want to eliminate the unknown ${\bf v}$ by using two (or more) instruments. From $R^{-1}(2 \tilde{\alpha}_i) {\bf u}_i = {\bf v}$, this gives us ${\bf u}_i = R\left(2(\tilde{\alpha}_i-\tilde{\alpha}_j)\right){\bf u}_j$ using the relationship $R(2 \tilde{\alpha}_i)R^{-1}(2 \tilde{\alpha}_j) = R\left(2(\tilde{\alpha}_i-\tilde{\alpha}_j)\right)$. This means that we can always find the difference $\tilde{\alpha}^-_{ij} = \alpha_i-\alpha_j=\alpha^-_{ij}$ as we have discussed in the main text. 

However, unfortunately there exists no such relationship for $\tilde{\alpha}^+_{ij}$ -- one cannot ``undo'' the rotation of $R(2 \tilde{\alpha}_i)$ by \emph{adding} another rotation to it since we do not know what the original datum point is. For example,  one can try adding the observation vectors 
\begin{eqnarray}
    \left(R(2\tilde{\alpha}_i)+R(2\tilde{\alpha}_j)\right){\bf v} &=& {\bf u}_i+{\bf u}_j~,\\
    \left(R(2\tilde{\alpha}_i)-R(2\tilde{\alpha}_j)\right){\bf v} &=& {\bf u}_i-{\bf u}_j~,
\end{eqnarray}
but solving for ${\bf v}$ yields the relationship
\begin{equation}
\twodsquaremat{0}{\cot(\tilde{\alpha}^-_{ij})}{-\cot(\tilde{\alpha}^-_{ij})}{0}({\bf u}_i-{\bf u}_j)={\bf u}_i+{\bf u}_j
\end{equation}

Again, $\tilde{\alpha}^+_{ij}$ is not present. In terms of correlation functions, this conundrum manifests itself as an identity as follows. Starting from the power spectra in \cref{correlation}, we can construct the following identities: 

\begin{equation}
     \begin{cases}
      \underbrace{C^{EE,o}_{\ell, \; ij}-C^{BB,o}_{\ell, \; ij}}_{A_{ij}}=(C^{EE}_{\ell}-C^{BB}_{\ell}) \cos(2 \tilde{\alpha}^+_{ij})-2C^{EB}_{\ell} \sin(2 \tilde{\alpha}^+_{ij}) \\ \underbrace{C^{EB,o}_{\ell, \; ij}+C^{EE,o}_{\ell, \; ji}}_{B_{ij}}= (C^{EE}_{\ell}-C^{BB}_{\ell}) \sin(2 \tilde{\alpha}^+_{ij}) +2C^{EB}_{\ell} \cos(2 \tilde{\alpha}^+_{ij})
    \end{cases}.
    \label{totaleq}
\end{equation}

We can recast this in a vector form : $X=  (A_{ij} , \, B_{ij})^\text{T}$ and $Y=( C^{EE}_{\ell}-C^{BB}_{\ell} , \, 2C^{EB}_{\ell} )^\text{T}$, and then it is clear that one is simply the rotation of the other: 
\begin{equation}
    X=R(2 \tilde{\alpha}_{ij}^+) Y,
\end{equation}
with $R$ the usual rotation matrix. Now, since $R$ is an element of the special orthogonal group $SO(2)$, this means that $R^2 =1$, i.e there is an identity that kills one equation. Because of that, we have no way of ever getting $\tilde{\alpha}_{ij}^+$ from the observables $A_{ij}$ and $B_{ij}$ only. Since this is proven at the level of the $E$ and $B$ fields, there is really no way out (i.e no clever combination, or higher order correlation) that will ever solve for $\tilde{\alpha}_{ij}^+$, therefore no way to get $\alpha_i$.

\section{Additional data}
The data used for \cref{fig:comparison_simulation} and \cref{birefringence_total} is detailed in \cref{tab:comparison_simulation}.
 and \cref{birefringence_LAT_Planck}
 
 \begin{table}[h!]
\centering
\begin{tabular}{|c|c|c|c|}
\hline
Experiment pair & $f$ [GHz] & Analytical [$^\circ$] & Simulation [$^\circ$] \\ 
\hline
SAT + \textit{Planck} & 93 / 100 & 0.28 & 0.30 \\
SAT + \textit{Planck} & 145 / 143 & 0.15 & 0.17 \\
SAT + \textit{Planck} & 225 / 217 & 0.25 & 0.28 \\
\hline
SAT + LAT & 93 & 0.06 & 0.07 \\
SAT + LAT & 145 & 0.07 & 0.08 \\
SAT + LAT & 225 & 0.14 & 0.18 \\
\hline
\end{tabular}
\caption{Comparison of the calibration precision $\sigma_{R_{ij}}$ between analytical estimates and simulations for SAT + \textit{Planck} and SAT + LAT frequency combinations. For SAT + \textit{Planck}, the frequencies are shown as SAT / \textit{Planck}.}
\label{tab:comparison_simulation}
\end{table}

\begin{table}[h!]
\centering
\begin{tabular}{|c|c|c|c|c|}
\hline
Experiment & $f$ [GHz] & $\sigma_\alpha$ [$^\circ$] & $\sigma_{\tilde{\alpha}}$ [$^\circ$] & $\sigma_\beta$ [$^\circ$] \\ 
\hline
LAT     & 93  & 0.10 & 0.011 & 0.17 \\ 
LAT     & 145 & 0.10 & 0.013 & 0.17 \\ 
LAT     & 225 & 0.16 & 0.036 & 0.28 \\ 
\hline
\textit{Planck}  & 100 & 0.28 & 0.11  & 0.50 \\ 
\textit{Planck}  & 143 & 0.17 & 0.043 & 0.30 \\ 
\textit{Planck}  & 217 & 0.27 & 0.095  & 0.48 \\ 
\hline
\end{tabular}
\caption{Comparison of the $1\sigma$ deviation on the isotropic birefringence angle $\beta$ for LAT and \textit{Planck} for different frequency channels.}
\label{birefringence_LAT_Planck}
\end{table}
\bibliography{main}
\end{document}